\documentclass{article}
\usepackage{spconf,amsmath,graphicx}
\usepackage{xcolor}
\usepackage{amssymb}
\usepackage{subcaption}
\usepackage{multirow}
\usepackage{multicol}
\usepackage{makecell}
\usepackage{soul}

\newif\ifdraft
\drafttrue

\ifdraft

\newcommand{\khremove}[1]{\textcolor{magenta}{{\st{#1}}}}

\else

\newcommand{\khremove}[1]{{}}
\fi

\title{SRU++: Pioneering Fast Recurrence with Attention for Speech Recognition}
\name{Jing Pan$^{\dagger}$, Tao Lei$^{\dagger}$, Kwangyoun Kim$^{\dagger}$, Kyu J. Han$^{\dagger}$, Shinji Watanabe$^{\ddagger}$}
\address{$^{\dagger}$ASAPP Inc., Mountain View, CA, USA \\
        $^{\ddagger}$Carnegie Mellon University, Pittsburgh, PA, USA
}

\begin{document}
%
\maketitle
\begin{abstract}
The Transformer architecture has been well adopted as a dominant  architecture in most sequence transduction tasks including automatic speech recognition (ASR), since its attention mechanism excels in capturing long-range dependencies. While models built solely upon attention can be better parallelized than regular RNN, a novel network architecture, SRU++, was recently proposed. By combining the fast recurrence and attention mechanism, SRU++ exhibits strong capability in sequence modeling and achieves near-state-of-the-art results in various language modeling and machine translation tasks with improved compute efficiency. In this work, we present the advantages of applying SRU++ in ASR tasks by comparing with Conformer across multiple ASR benchmarks and study how the benefits can be generalized to long-form speech inputs. On the popular LibriSpeech benchmark, our SRU++ model achieves 2.0\% / 4.7\% WER on test-clean / test-other, showing competitive performances compared with the state-of-the-art Conformer encoder under the same set-up. Specifically, SRU++ can surpass Conformer on long-form speech input with a large margin, based on our analysis.

\end{abstract}
\section{Introduction}
\label{sec:intro}

Since deep neural networks (DNNs) were applied in automatic speech recognition (ASR) tasks, various DNN architectures which helped boosting the ASR accuracy have been proposed. Recurrent neural networks (RNNs) have been widely used given that they can effectively capture the temporal dependencies of a speech frame sequence \cite{graves2013speech,chiu2018state}. Convolutional neural networks (CNNs), originally popular in the computer vision domain, have also become preferred options for ASR tasks. By carefully tuning the receptive fields, residual connections and temporal resolution of the convolutional layers, deep CNNs can yield very competitive results \cite{han2020contextnet,han2021multistream}. 

However, it is well known that both RNN and CNN struggle to model long sequences.
Typical RNN variants such as LSTM and GRU utilize gating operations to alleviate gradient explosion/vanishing problems, but this also leads to catastrophic forgetting, where information carried by hidden states are overwritten.
On the other hand, CNN uses local connectivity and therefore requires many more layers to gain enough perceptive field for global context, which also creates difficulties for optimization.


Recently, the Transformer architecture with self-attention \cite{vaswani2017attention} is trending in ASR tasks due to its dominant performance in modeling the long-range dependencies
\cite{wang2020transformer,moritz2020streaming}.
The architecture strikes a great balance between model capacity and training efficiency; (a) attention permits learning dependencies between any pair of time steps in a given sequence, (b) the associated computation of the attention mechanism can be formatted in a couple of matrix multiplications and are therefore highly parallelizable, and (c) scaling model capacity can be achieved by increasing the attention dimension and the number of attention heads, aside from layer stacking. 

Despite the great success of Transformer, several research have found that the powerful attention mechanism can still be complemented by traditional neural components that are well-suited for capturing fine-grain local context \cite{merity2019single,lei2021attention}.
For example, Conformer \cite{gulati2020conformer} proposes adding a convolution module leveraging gating mechanisms in point-wise convolution and gated linear unit (GLU) on top of multi-head self-attention to augment Transformer, yielding faster convergence and state-of-the-art performance on various speech recognition datasets. 
If convolutions were effective at enhancing the performance of Transformer models by complementing the attention module, could we achieve similar or better synergy by taking the advantage of the recent development of recurrent networks for attention?

In this work, we present the application of a novel network architecture called SRU++ introduced by \cite{lei2021attention} in ASR task, which is, to our best knowledge, the first attempt to apply such an attention-recurrence fusion network in speech recognition. The basic idea of SRU++ is to leverage a highly parallelizable recurrent neural network structure called SRU \cite{lei2017simple} and perform attention before the recurrence. We conduct ASR experiments on three popular public datasets and compare the results with the Transformer and Conformer encoders implemented and open-sourced in \cite{guo2021recent}. Results show that our SRU++ encoder model can perform on par with the state-of-the-art Conformer model under the same set-up. 
In addition, our analysis shows that SRU++ generalizes exceptionally well on long-form speech, outperforming the Conformer model by a large margin.

\section{SRU++ Encoder Model}
\label{sec:sruppencoder}
\subsection{SRU}
We start by describing simple recurrent unit (SRU)~\cite{lei2017simple} to give necessary context of our proposed ASR model.
An SRU layer contains the following operations:
\begin{equation}
\begin{split}
& \mathbf f[t] = \sigma(\mathbf{Wx}[t] + \mathbf{v} \odot \mathbf{c}[t-1] + \mathbf{b}) \\
& \mathbf{r}[t] = \sigma(\mathbf{W'x}[t] + \mathbf{v'} \odot \mathbf{c}[t-1] + \mathbf{b'}) \\
& \mathbf{c}[t] = \mathbf{f}[t] \odot \mathbf{c}[t-1] + (1-\mathbf{f}[t]) \odot (\mathbf{W''x}[t]) \\
& \mathbf{h}[t] = \mathbf{r}[t] \odot \mathbf{c}[t] + (1-\mathbf{r}[t]) \odot \mathbf{x}[t]
\end{split}
\label{eqSRU}
\end{equation}
Where $\odot$ is element-wise multiplication, $\mathbf W$, $\mathbf W'$ and $\mathbf W''$ are weight matrices, and $\mathbf v$, $\mathbf v'$, $\mathbf b$ and $\mathbf b'$ are also trainable vectors. Similar to LSTM and GRU, SRU uses gating operations such as the forget gate $\mathbf{f}[t]$ and reset gate $\mathbf{r}[t]$ to control the information flow from the internal memory $\mathbf{c}[t-1]$ and current input $\mathbf{x}[t]$. 
In order to enhance parallelism, multiplications for the three matrices are combined and computed as a single operation. 
Given the input sequence $\mathbf{X} = \{\mathbf{x}[1], \dots, \mathbf{x}[L]\}$ where $\mathbf{x}[t] \in \mathbb{R}^d$ is a $d$-dimensional vector, the matrix multiplication can be written as:
\begin{equation}
    \mathbf{U} = \begin{bmatrix} \mathbf{W} \\ \mathbf{W'} \\ \mathbf{W''} \end{bmatrix} \mathbf{X^\top}
\label{eqU}
\end{equation}
where $\mathbf{U} \in \mathbb{R}^{L\times 3 \times d}$ is the output tensor. Then the SRU operations above can be rewritten into:
\begin{equation}
\begin{split}
& \mathbf f[t] = \sigma(\mathbf{U}[t,0] + \mathbf{v} \odot \mathbf{c}[t-1] + \mathbf{b}) \\
& \mathbf{r}[t] = \sigma(\mathbf{U}[t,1] + \mathbf{v'} \odot \mathbf{c}[t-1] + \mathbf{b'}) \\
& \mathbf{c}[t] = \mathbf{f}[t] \odot \mathbf{c}[t-1] + (1-\mathbf{f}[t]) \odot \mathbf{U}[t,2] \\
& \mathbf{h}[t] = \mathbf{r}[t] \odot \mathbf{c}[t] + (1-\mathbf{r}[t]) \odot \mathbf{x}[t]
\end{split}
\label{eqSRU_U}
\end{equation}
As shown in the equations, each dimension of the hidden vectors is independent and all operations are parallelizable, element-wise computation.
These operations are computed using an optimized SRU CUDA and/or CPU kernel.

\subsection{SRU++}
\begin{figure}[t!]
    \centering
    \begin{subfigure}[t]{0.45\linewidth}
        \centering
        \includegraphics[height=1.6in]{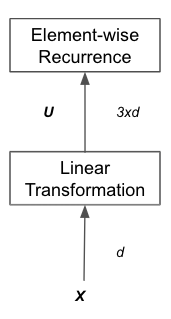}
        \caption{SRU}
    \end{subfigure}%
    ~ 
    \begin{subfigure}[t]{0.55\linewidth}
        \centering
        \includegraphics[height=2.0in]{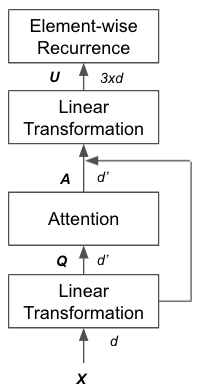}
        \caption{SRU++}
    \end{subfigure}
    \caption{Comparing structures of SRU and SRU++ layer.}
\label{sruppstructure}
\end{figure}

SRU++ is a modified version of SRU which incorporates attention operations into the recurrence. 
Specifically, the computation of $\mathbf{U}$ (Eq. \ref{eqU}), which is a linear transformation, can be alternatively formulated into a self-attention operation and enhance the model capacity. Given the input sequence $\mathbf{X} \in \mathbb{R}^{L \times d} $, the major attention components query, key and value vectors are computed by
\begin{equation}
\begin{split}
   & \mathbf{Q} = \mathbf{W^q}\mathbf{X^\top} \\
   & \mathbf{K} = \mathbf{W^{k}Q} \\
   & \mathbf{V} = \mathbf{W^{v}Q}
\end{split}
\label{eqQKV}
\end{equation}
where $\mathbf{W^q} \in \mathbb{R}^{d' \times d}$, $\mathbf{W^k} \in \mathbb{R}^{d' \times d'}$, $\mathbf{W^v} \in \mathbb{R}^{d' \times d'}$ are trainable projection matrices, $d'$ is an attention dimension which is typically smaller than $d$. To save some parameters, $\mathbf{K}$ and $\mathbf{V}$ are derived from $\mathbf{Q}$. Next, we can compute the attention output $\mathbf{A} \in \mathbb{R}^{d'\times L}$ with
\begin{equation}
\mathbf{A}^\top = softmax(\frac{\mathbf{Q}^\top\mathbf{K}}{\sqrt{d'}})\mathbf{V^\top}
\label{eqAT}
\end{equation}
With this attention computation, Eq. \ref{eqU} can now be rewritten as
\begin{equation}
\mathbf{U}^\top = \mathbf{W^o}(\mathbf{Q} + \alpha \cdot \mathbf{A})
\label{eqUT}
\end{equation}
where $\mathbf{W^o}\in \mathbb{R}^{3d \times d'}$ is parameterized and $\alpha$ is a trainable scalar. ($\mathbf{Q} + \alpha \cdot \mathbf{A}$) denotes a residual connection which improves the gradient propagation and therefore stabilizes the training. Fig \ref{sruppstructure} demonstrates the structure of SRU++ in comparison with SRU.

\begin{figure*}[t!]
    \centering
    \includegraphics[height=0.55in, width=0.8\linewidth]{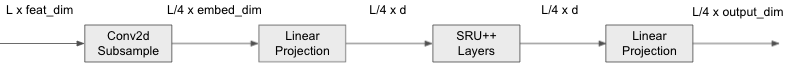}
    \caption{Overview of SRU++ encoder.}
    \label{fig:srupp_encoder}
    \vspace{-0.35cm}
\end{figure*}

\subsection{SRU++ Encoder}
Our SRU++ encoder model starts with a 2D convolution based subsampling layers which subsamples and shortens the input speech feature sequence, followed by a linear layer to project the output to the recurrence hidden size ($d$). Then the output is fed into a list of SRU++ layers. Depending on the expected model output dimension, there is another optional linear layer to transform the SRU++ outputs to the expected dimension. Fig \ref{fig:srupp_encoder} visualizes the SRU++ encoder architecture. Note that $L$ denotes the input sequence length, $feat\_dim$ is the feature dimension, $embed\_dim$ is the dimension of the subsampled output, and $output\_dim$ is the encoder output dimension.
\section{Data and experimental setup}
\label{sec:pagestyle}
\vspace{-0.2cm}
ASR experiments are conducted on 3 public ASR data sets. LibriSpeech \cite{panayotov2015LibriSpeech} contains approximately 1000 hours of 16kHz English audio-book recordings, among which 960 hours are used for training partition. AISHELL-1 \cite{bu2017aishell} has 178 hours of 16kHz Chinese Mandarin recordings with various accents. TEDLIUM3 \cite{hernandez2018ted} includes 456 hours of 16kHz English TED talks.

We leverage the Transformer and Conformer models trained and open-sourced by \cite{guo2021recent} as the baseline models. To make sure the results are comparable, we follow the exact same data preparation and feature extraction configurations. We implement our SRU++ encoder in ESPNet \cite{watanabe2018espnet} framework. Besides, we have the identical configurations for decoder structures, Conv2d subsampling layers and text tokenizers compared with the baseline models. By scaling the SRU++ recurrence hidden size ($d$) and attention dimension ($d'$), we can control the model parameter sizes at the same scale as the corresponding baseline models. We turn on bidirectional recurrence for all of the SRU++ models for better model capacity. In such case, the hidden size $d$ would be divided by 2 for each direction. We train our models using the same effective batch sizes as the baseline models. In \cite{guo2021recent}, character/token/word-level neural language models are used for shallow fusion during the decoding, we apply the exact same language models respectively. Experiments in \cite{lei2021attention} show that more attention heads do not improve performance of SRU++ and no positional encoding is required. Therefore, in this work, we assign only 1 head to all of the SRU++ layers and disable positional encoding for SRU++.

We adopt the same training and decoding mechanism. Data augmentation techniques such as SpecAugment \cite{park2019specaugment} and speed perturbation with scaling factors of 0.9, 1.0 and 1.1 are applied. In the same way as the baseline models, we employ the same CTC-seq2seq joint training and perform CTC-seq2seq joint decoding in the test phase for better performance. Data preparation and the Transformer decoder configs can be found in the ESPNet recipes\footnote{e.g. https://github.com/espnet/espnet/tree/master/egs2/librispeech/asr1}. The hyper-parameters such as learning rate and weight decay are tuned on the corresponding dev sets. We list our set-ups in Table \ref{table:params}.

\begin{table}[ht]
\centering
\renewcommand{\arraystretch}{1.1}
\begin{tabular}{ccccc}
\Xhline{3\arrayrulewidth}
\textbf{Dataset} & \textbf{$d$} & \textbf{$d'$} & $lr$ & $wd$ \\

\hline
\hline
LibriSpeech & 3328 & 416 & 7e-4 & 0.05\\
\hline
AISHELL-1 & 2176 & 272 & 2.5e-4 & 0.05\\
\hline
TEDLIUM3 & 2176 & 272 & 4.4e-4 & 0.05\\
\Xhline{3\arrayrulewidth}
\end{tabular}
\caption{Parameters of the selected models}
\label{table:params}
\vspace{-0.35cm}
\end{table}

\begin{table*}[ht]
\centering
\renewcommand{\arraystretch}{1.3}
\begin{tabular}{cccccc}
\Xhline{3\arrayrulewidth}
\multicolumn{1}{c}{\textbf{Dataset}} & \multicolumn{1}{c}{\textbf{Metric}} & \multicolumn{1}{c}{\textbf{Evaluation Sets}} & \multicolumn{1}{c}{\textbf{Transformer}} & \multicolumn{1}{c}{\textbf{Conformer}} & \multicolumn{1}{c}{\textbf{SRU++}} \\

\cline{2-5}
\hline
\hline
LibriSpeech & WER & \{dev, test\}\_\{clean,other\} & 2.1 / 5.3 / 2.5 / 5.5 & 1.9 / \textbf{4.6} / 2.1 / 4.7 & \textbf{1.8} / 4.8 / \textbf{2.0} / 4.7 \\
\hline
AISHELL-1 & CER & dev / test & 6.0 / 6.7 & 4.4 / 4.7 & 4.4 / 4.7 \\
\hline
TEDLIUM3 & WER & dev / test & 10.8 / 8.4 & 9.6 / \textbf{7.6} & \textbf{8.3} / 8.0\\
\Xhline{3\arrayrulewidth}
\end{tabular}
\vspace{0.5em}
\caption{Error rates (in \%) comparison among ASR models with Transformer, Conformer and SRU++ encoders.}
\label{table:wer_compare}
\end{table*}

\section{Results and analysis}
\label{sec:results}
\subsection{Accuracy Benchmark}
Table \ref{table:wer_compare} compares the WER/CER of the ASR models with Transformer, Conformer and SRU++ encoder respectively. Note that results of the Transformer and Conformer are taken from \cite{guo2021recent}, the corresponding trained models along with the language models are also available on ESPNet toolkit. It can be told from the table that under the same experimental set-up, the SRU++ consistently outperforms Transformer as Conformer does. 
In the meantime, SRU++ scores competitively in WER/CER against Conformer. The highlight is the 5\% relative improvement on LibriSpeech test-clean and ~14\% relative improvement on TEDLIUM3 dev set. On the other hand, we can also observe that Conformer performs better on some evaluation sets such as LibriSpeech dev-other and TEDLIUM3 test set. Notably, in ASR task, SRU++ with single attention head exhibits similar model capacity as Conformer. It is evident that the recurrent connection can provide strong support for sequence modeling, resulting in the reduction of dependencies on attention modules.

\subsection{Performance On Long-form Speech}
\begin{figure}[t!]
    \vspace{-0.6cm}
    \centering
    \includegraphics[height=2.7in, width=\linewidth]{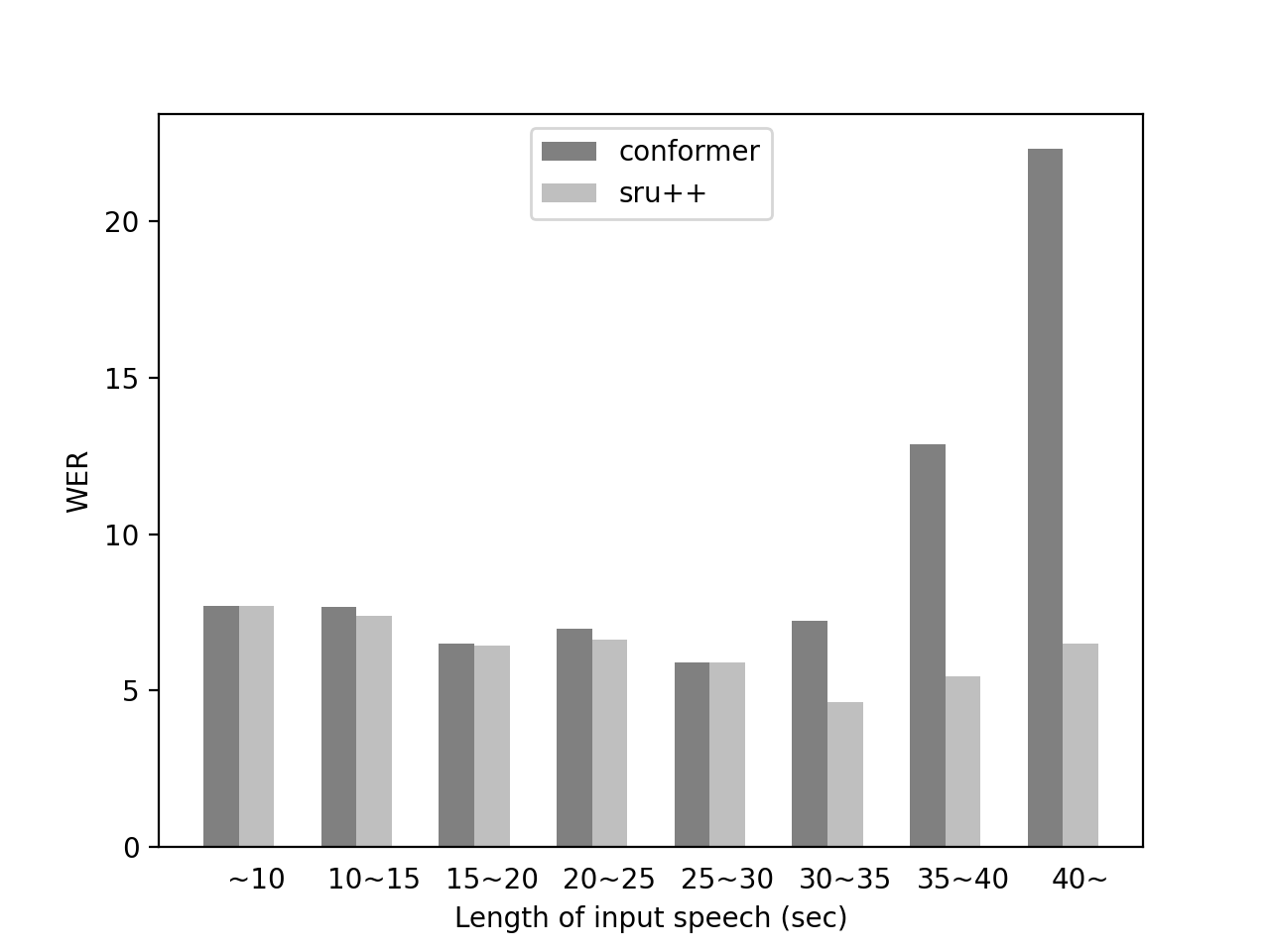}
    \caption{Avg. WER (\%) distribution by speech length (sec).}
    \label{fig:decode_long}
    \vspace{0.5cm}
\end{figure}

A potential advantage of SRU++ encoder is that it is trained without positional encoding, since recurrence naturally injects the positional information. Consequently, SRU++ might be better generalized to the longer speech sequences which rarely exist in the training data. To test this hypothesis, we concatenate every 3 utterances in the same book chapter in LibriSpeech test-clean / test-other datasets and aggregate the concatenated utterances into a new test set with avg. audio length longer than 20 sec. Fig \ref{fig:decode_long} shows the WER distribution per audio length. The performance of Conformer and SRU++ is similar when length is under 30 seconds. Conformer degrades drastically for audio longer than 30 sec while SRU++ stays relatively stable. For the audio longer than 40 seconds, WER of Conformer is 4 times of  SRU++. Therefore, we can conclude that SRU++ is more robust on the long-form speech than Conformer.

\subsection{Compute Efficiency}
We measure the number of floating point operations (in GFlops) during forward calls for the Conformer and SRU++ encoders. The GFlops is profiled using DeepSpeed\footnote{https://github.com/microsoft/DeepSpeed} toolkit. Note that two models are trained on LibriSpeech and contain similar amount of parameters. According to Table \ref{table:flops_compare}, SRU++ yields smaller GFlops, indicating lighter computation. Thanks to the well-parallelized element-wise recurrence in SRU, SRU++ can maintain a good balance of model capacity and compute efficiency.

\begin{table}[!t]
\centering
\renewcommand{\arraystretch}{1.1}
\begin{tabular}{ccc}
\Xhline{3\arrayrulewidth}
\textbf{Model} & \textbf{GFlops} & \textbf{\#Params(million)} \\
\hline
\hline
Conformer & 64.3 & 83.2 \\
\hline
SRU++ & 62.0 & 81.7 \\
\Xhline{3\arrayrulewidth}
\end{tabular}
\vspace{0.5em}
\caption{GFlops comparison between Conformer and SRU++, when the length of input sequence is 1000.}
\label{table:flops_compare}
\end{table}

\subsection{Unidirectional vs. Bidirectional}
\begin{figure}[htb]
\begin{minipage}[b]{0.3\linewidth}
  \centering
  \centerline{\includegraphics[width=3cm]{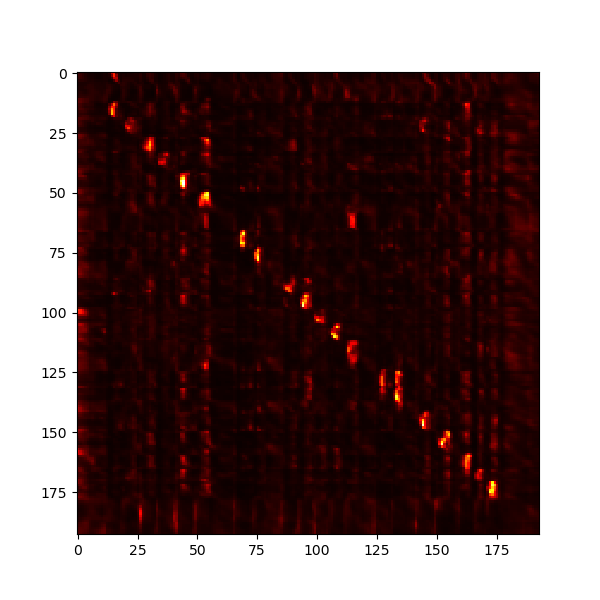}}
  \centerline{(a) Conformer}\medskip
\end{minipage}
\hfill
\begin{minipage}[b]{0.3\linewidth}
  \centering
  \centerline{\includegraphics[width=3cm]{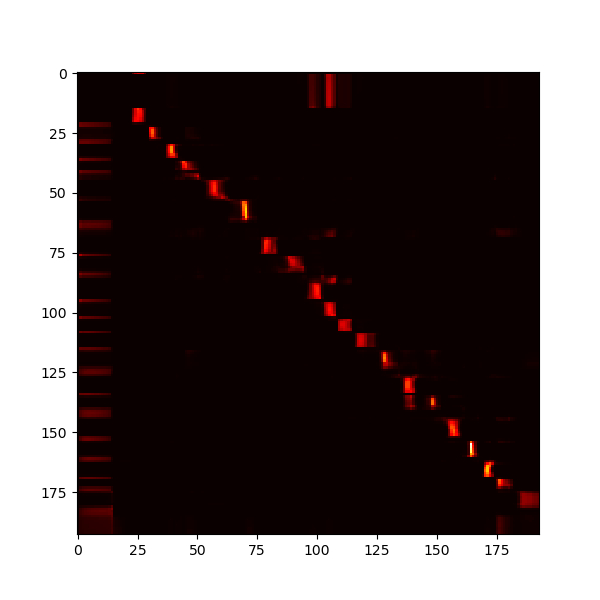}}
  \centerline{(b) Uni-SRU++}\medskip
\end{minipage}
\hfill
\begin{minipage}[b]{0.3\linewidth}
  \centering
  \centerline{\includegraphics[width=3cm]{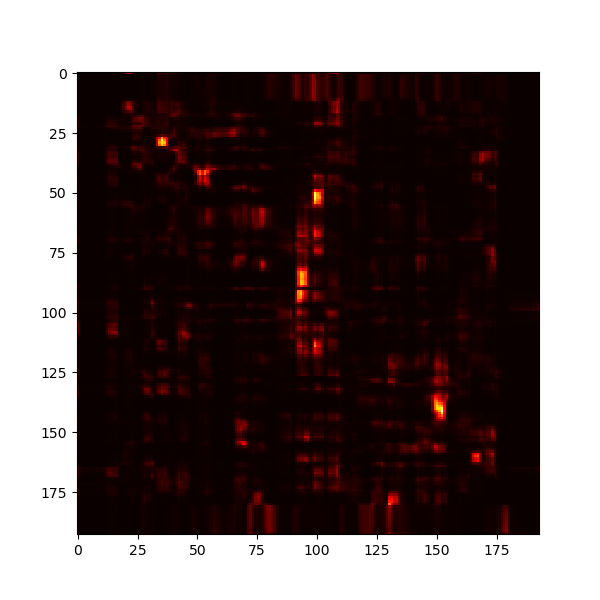}}
  \centerline{(c) Bi-SRU++}\medskip

\end{minipage}
\caption{Average attention weights from the last encoder layer.}
\label{fig:attn_weight}
  \vspace{-0.5cm}
\end{figure}

We study the impact of bidirectional recurrence using AISHE\\LL-1. We trained another model using unidirectional SRU++ encoder with the same $d$ and $d'$. The CERs of the two models are listed in Table \ref{table:bivsuni}. Disabling the backward recurrence results in about 10\% CER degradation. We feed a sequence to the model and extract the averaged attention weight matrices over attention heads from the last encoder layer and visualize them in Fig \ref{fig:attn_weight}. We observe more diagonal patterns for Conformer and unidirectional SRU++. 
Interestingly, we do not observe a strong diagonal patterns in the attention heat map of bidirectional SRU++.
We hypothesize that for speech input, the most important information of acoustic features remains local, therefore the attention weights tend to be diagonal.
However, since bidirectional recurrence in SRU++ is already strong and flexible at capturing local context, the attention has the most freedom to focus on non-trivial and long-range dependencies.
\begin{table}[ht]
\centering
\renewcommand{\arraystretch}{1.1}
\begin{tabular}{ccc}
\Xhline{3\arrayrulewidth}
\textbf{Config} & \textbf{dev} & \textbf{test} \\

\hline
\hline
Unidirectional & 4.9 & 5.1 \\
\hline
Bidirectional & 4.4 & 4.7 \\
\Xhline{3\arrayrulewidth}
\end{tabular}
\vspace{0.5em}
\caption{CER(\%) on AISHELL-1}
\label{table:bivsuni}
  \vspace{-0.5cm}
\end{table}

\section{Conclusion}
\label{sec:conclusion}

In this work, we present the ASR application of SRU++, an architecture that incorporates attention with fast recurrence. We study the robustness of the SRU++ based model for long-form speech and demonstrate the importance of the fast recurrence. The SRU++ encoder model exhibits competitive results on multiple public data sets with less compute cost compared with the state-of-the-art Conformer model. In the future, we will explore SRU++ as a decoder, and the potential of SRU++ applied to self-supervised pre-training tasks.

\bibliographystyle{IEEEbib}
\bibliography{refs}

\begin{thebibliography}{10}

\bibitem{graves2013speech}
Alex Graves, Abdel-rahman Mohamed, and Geoffrey Hinton,
\newblock ``Speech recognition with deep recurrent neural networks,''
\newblock in {\em 2013 IEEE international conference on acoustics, speech and
  signal processing}. Ieee, 2013, pp. 6645--6649.

\bibitem{chiu2018state}
Chung-Cheng Chiu, Tara~N Sainath, Yonghui Wu, Rohit Prabhavalkar, Patrick
  Nguyen, Zhifeng Chen, Anjuli Kannan, Ron~J Weiss, Kanishka Rao, Ekaterina
  Gonina, et~al.,
\newblock ``State-of-the-art speech recognition with sequence-to-sequence
  models,''
\newblock in {\em 2018 IEEE International Conference on Acoustics, Speech and
  Signal Processing (ICASSP)}. IEEE, 2018, pp. 4774--4778.

\bibitem{han2020contextnet}
Wei Han, Zhengdong Zhang, Yu~Zhang, Jiahui Yu, Chung-Cheng Chiu, James Qin,
  Anmol Gulati, Ruoming Pang, and Yonghui Wu,
\newblock ``Contextnet: Improving convolutional neural networks for automatic
  speech recognition with global context,''
\newblock {\em arXiv preprint arXiv:2005.03191}, 2020.

\bibitem{han2021multistream}
Kyu~J Han, Jing Pan, Venkata Krishna~Naveen Tadala, Tao Ma, and Dan Povey,
\newblock ``Multistream cnn for robust acoustic modeling,''
\newblock in {\em ICASSP 2021-2021 IEEE International Conference on Acoustics,
  Speech and Signal Processing (ICASSP)}. IEEE, 2021, pp. 6873--6877.

\bibitem{vaswani2017attention}
Ashish Vaswani, Noam Shazeer, Niki Parmar, Jakob Uszkoreit, Llion Jones,
  Aidan~N Gomez, {\L}ukasz Kaiser, and Illia Polosukhin,
\newblock ``Attention is all you need,''
\newblock in {\em Advances in neural information processing systems}, 2017, pp.
  5998--6008.

\bibitem{wang2020transformer}
Yongqiang Wang, Abdelrahman Mohamed, Due Le, Chunxi Liu, Alex Xiao, Jay
  Mahadeokar, Hongzhao Huang, Andros Tjandra, Xiaohui Zhang, Frank Zhang,
  et~al.,
\newblock ``Transformer-based acoustic modeling for hybrid speech
  recognition,''
\newblock in {\em ICASSP 2020-2020 IEEE International Conference on Acoustics,
  Speech and Signal Processing (ICASSP)}. IEEE, 2020, pp. 6874--6878.

\bibitem{moritz2020streaming}
Niko Moritz, Takaaki Hori, and Jonathan Le,
\newblock ``Streaming automatic speech recognition with the transformer
  model,''
\newblock in {\em ICASSP 2020-2020 IEEE International Conference on Acoustics,
  Speech and Signal Processing (ICASSP)}. IEEE, 2020, pp. 6074--6078.

\bibitem{merity2019single}
Stephen Merity,
\newblock ``Single headed attention rnn: Stop thinking with your head,''
\newblock {\em arXiv preprint arXiv:1911.11423}, 2019.

\bibitem{lei2021attention}
Tao Lei,
\newblock ``When attention meets fast recurrence: Training language models with
  reduced compute,''
\newblock in {\em Proceedings of the 2021 Conference on Empirical Methods in
  Natural Language Processing (EMNLP)}, 2021.

\bibitem{gulati2020conformer}
Anmol Gulati, James Qin, Chung-Cheng Chiu, Niki Parmar, Yu~Zhang, Jiahui Yu,
  Wei Han, Shibo Wang, Zhengdong Zhang, Yonghui Wu, et~al.,
\newblock ``Conformer: Convolution-augmented transformer for speech
  recognition,''
\newblock {\em arXiv preprint arXiv:2005.08100}, 2020.

\bibitem{lei2017simple}
Tao Lei, Yu~Zhang, Sida~I Wang, Hui Dai, and Yoav Artzi,
\newblock ``Simple recurrent units for highly parallelizable recurrence,''
\newblock in {\em Proceedings of the 2018 Conference on Empirical Methods in
  Natural Language Processing (EMNLP)}, 2018.

\bibitem{guo2021recent}
Pengcheng Guo, Florian Boyer, Xuankai Chang, Tomoki Hayashi, Yosuke Higuchi,
  Hirofumi Inaguma, Naoyuki Kamo, Chenda Li, Daniel Garcia-Romero, Jiatong Shi,
  et~al.,
\newblock ``Recent developments on espnet toolkit boosted by conformer,''
\newblock in {\em ICASSP 2021-2021 IEEE International Conference on Acoustics,
  Speech and Signal Processing (ICASSP)}. IEEE, 2021, pp. 5874--5878.

\bibitem{panayotov2015LibriSpeech}
Vassil Panayotov, Guoguo Chen, Daniel Povey, and Sanjeev Khudanpur,
\newblock ``Librispeech: an asr corpus based on public domain audio books,''
\newblock in {\em 2015 IEEE international conference on acoustics, speech and
  signal processing (ICASSP)}. IEEE, 2015, pp. 5206--5210.

\bibitem{bu2017aishell}
Hui Bu, Jiayu Du, Xingyu Na, Bengu Wu, and Hao Zheng,
\newblock ``Aishell-1: An open-source mandarin speech corpus and a speech
  recognition baseline,''
\newblock in {\em 2017 20th Conference of the Oriental Chapter of the
  International Coordinating Committee on Speech Databases and Speech I/O
  Systems and Assessment (O-COCOSDA)}. IEEE, 2017, pp. 1--5.

\bibitem{hernandez2018ted}
Fran{\c{c}}ois Hernandez, Vincent Nguyen, Sahar Ghannay, Natalia Tomashenko,
  and Yannick Esteve,
\newblock ``Ted-lium 3: twice as much data and corpus repartition for
  experiments on speaker adaptation,''
\newblock in {\em International conference on speech and computer}. Springer,
  2018, pp. 198--208.

\bibitem{watanabe2018espnet}
Shinji Watanabe, Takaaki Hori, Shigeki Karita, Tomoki Hayashi, Jiro Nishitoba,
  Yuya Unno, Nelson Enrique~Yalta Soplin, Jahn Heymann, Matthew Wiesner, Nanxin
  Chen, et~al.,
\newblock ``Espnet: End-to-end speech processing toolkit,''
\newblock {\em arXiv preprint arXiv:1804.00015}, 2018.

\bibitem{park2019specaugment}
Daniel~S Park, William Chan, Yu~Zhang, Chung-Cheng Chiu, Barret Zoph, Ekin~D
  Cubuk, and Quoc~V Le,
\newblock ``Specaugment: A simple data augmentation method for automatic speech
  recognition,''
\newblock {\em arXiv preprint arXiv:1904.08779}, 2019.

\end{thebibliography}

\end{document}